\documentclass{aa}  

\usepackage{graphicx}
\usepackage{subcaption}
\usepackage{color}
\usepackage{soul}
\usepackage{siunitx}
\DeclareSIUnit{\AU}{\text{au}}
\usepackage[hidelinks, breaklinks=true]{hyperref}
\usepackage{txfonts}

\begin{document}

   \title{Phase angle dependency of the dust cross section in a cometary coma}

   \subtitle{A simple numerical model}

   \author{F. Keiser
          \and
          J. Markkanen
          \and
          J. Agarwal
          }

   \institute{Institut für Geophysik und extraterrestrische Physik (IGEP), Technische Universität Braunschweig,
   \email{f.keiser@tu-bs.de}}

   \date{\today}

 
  \abstract
   {Rosetta/OSIRIS took optical measurements of the intensity of scattered light from the coma of 67P/Churyumov-Gerasimenko over a wide range of phase angles. These data have been used to measure the phase angle dependent radiance profile of the dust coma.}
   {We want to provide information about the column area densities of the dust coma as seen from Rosetta. This information in combination with the measured OSIRIS phase function can then be used to determine the scattering phase function of the dust particles.}
   {We use a simple numerical model to calculate the dust density in the coma. For this we neglect all forces but solar gravitation and radiation pressure. As this cannot describe particles close to the surface of the comet, we assume starting conditions at a sufficient distance. We evaluate the column area density as observed from Rosetta/OSIRIS and compare the results for different spacecraft positions, dust sizes and surface activity distributions.}
   {We find the phase angle dependence of the column area density to be largely independent of particle size and spacecraft positions. The determining factor is the activity distribution across the surface, especially the activity on the night side. For models with no night side activity, we find the column area density at high phase angles to be roughly two orders of magnitude larger than at low phase angles.}
   {The radiance profile measured from inside a cometary coma results from the combined effects of a phase angle dependent column area density and the scattering phase function. The radiance profile is therefore strongly dependent on the surface activity distribution, and -- unless the dust emission is isotropic -- any attempt to infer particle properties (as expressed through the scattering phase function) from such data must take into account and de-bias for this spatial variation of the dust column area density.}


   \maketitle
%

\section{Introduction}

The composition of comets has been of great interest for a long time, yet we still know relatively little about it, mainly due to the difficulty of obtaining measured data. The ESA mission Rosetta was the first mission to conduct measurements inside a cometary coma for an extended period of time \citep{Glassmeier2007, Taylor2017}. One tool was the onboard camera system \textsc{OSIRIS}, which was mainly used for observing the nucleus as well as the dust coma of the target of Rosetta, 67P/Churyumov-Gerasimenko  \citep{Keller2007}.

\citet{Bertini2017} use these measurements to determine the radiance of the coma of 67P/Churyumov-Gerasimenko as a function of phase angle using twelve \textsc{OSIRIS} image series taken between 29.03.2015 and 18.02.2016 at varying heliocentric and nucleocentric distances. After removing stray light effects, they find a u-shaped phase function, that is overall constant in time.

To interpret the u-shaped phase function numerical modelling and laboratory experiments of light scattering by single particles have been conducted. \citet{Moreno2018} showed numerically that large aligned dust particles can reproduce the peculiar u-shape, which was later confirmed by the laboratory experiments carried out by \cite{Munoz2020}. However,  \citet{Moreno2021} showed that the combined effect of gas drag and radiation forces and torques is not able to explain the required particle alignment.  In a different approach, a model using randomly oriented particles with strongly constrained internal structure was shown by \citet{Markkanen2018} to produce a u-shaped scattering phase function. Also, \citet{ACLR2019} showed experimentally that large dust particles can reproduce the u-shaped phase function but with a shallower minimum. Overall, it is difficult to reproduce the Rosetta/OSIRIS phase functions using numerical models or laboratory experiments. This may indicate that the Rosetta/OSIRIS radiance phase function does not represent the single scattering phase function of dust in the near coma due to the phase angle dependent column area density (defined below, see \autoref{eq:colareadens}).

In order to correlate the measured radiance profile of the dust coma to the properties of the dust particles, the density distribution of dust particles in the coma has to be known. In this paper we propose a simple numerical model to simulate this density. Then we investigate the column area density as seen from Rosetta in geometries similar to the measurements used by \citet{Bertini2017}. We investigate whether the exact observational geometry is of importance for the observed results and perform parameter studies over some poorly constrained variables, like the dust size and the activity distribution across the surface of 67P/Churyumov-Gerasimenko.

\section{Model}
\subsection{Acting forces}
\label{sec:forces}

The used physical model treats both the comet and dust particles as perfect spheres with homogeneous bulk density. In the context of this model we assume both densities to be identical ($\SI{500}{\kilo\gram\per\meter\cubed}$). The dominant forces are the gravitational pull of the sun
\begin{equation}
	F_G = \frac{G M_\odot}{r_h^2} \frac{4\rho\pi a^3}{3},
\end{equation}
and the radiation pressure of the sun
\begin{equation}
	F_r = \frac{Q_{pr}}{c} \frac{L_\odot}{4\pi r_h^2} \pi a^2,
\end{equation}
with the bulk density of the particle $\rho$, particle radius $a$, gravitational constant $G$, heliocentric distance $r_h$, solar luminosity $L_\odot$ and mass ${M_\odot}$, vacuum speed of light $c$ and the radiation pressure coefficient $Q_{pr}$. The latter is assumed to be $Q_{pr} = 1$ for particles larger than the wavelength of the observed light \citep{bohren-huffman1983}.

As these two forces are always acting in opposite directions (towards and away from the sun, respectively), the resulting force can be written as
\begin{equation}
    F_{net} = \left(1-\beta\right) F_G
    \label{eq:forces_0}
\end{equation}
with
\begin{equation}
	\beta = \frac{3L_\odot Q_{pr}}{16\pi G M_\odot c \rho a} = \SI{5.77e-4}{kg\per\meter\squared} \cdot\frac{Q_{pr}}{\rho a}.
\end{equation}

The radial acceleration experienced by any given particle is thus
\begin{equation}
    \ddot{r} = \frac{F_{net}}{m} = \left(1-\beta\right) \frac{G M_\odot}{r_h^2}.
    \label{eq:forces}
\end{equation}
The only dependence on the particle parameters is contained in $\beta$. All other acting forces are not taken into account, most notably the gravitational force of the comet as well as interactions between the dust particles and the gas in the coma and with each other. The gas drag can be neglected given a sufficient distance ($\approx \SI{e4}{\meter}$) between comet and dust particles \citep{Marschall2020, Gerig2018, Zakharov2018}.
The Hill radius for our model configuration is $r_H \approx \SI{5.7e6}{\meter}$. Even though we do carry out calculations for particles closer to the comet, we argue that the effect of nucleus gravity can be neglected as the initial speed of all particles is larger than the escape velocity of $v_{esc} \approx \SI{0.87}{\meter\per\second}$ (min. initial velocity $\SI{3}{\meter\per\second}$ see below).

We hence use the same equations of motion for the nucleus and the dust particles, the only difference being the value of $\beta$, with $\beta_{\rm 67P} << \beta_{\rm dust}$. For this model, we use a comet radius of $r_N = \SI{3e3}{\meter}$.

\subsection{Coordinate systems and initial conditions}
\label{sec:geom}

The intention of our simulations is to qualitatively investigate the relevance of a non-isotropic dust coma for the interpretation of the radiance profiles measured by \cite{Bertini2017}, and -- more general -- for the column area density profiles as seen from inside a cometary coma. We will not carry out simulations for every single epoch of the data analysed by \cite{Bertini2017}. Hence we use the following definitions and assumptions for our model:
\begin{enumerate}
    \item[D1.]{We carry out the simulations in a Cartesian coordinate system centred on the sun. The $\xi$-axis is aligned with the comet's heliocentric position vector at the time $t_0$. The $\eta$-axis lies in the orbital plane of the comet, is perpendicular to the $\xi$-axis and points to the same hemicycle as the comet's orbital velocity. The $\zeta$-axis is perpendicular to both. This coordinate system remains fixed in space throughout our simulations and is therefore an inertial system.}
    \item[D2.] The acceleration is parallel to the radial direction with $\ddot{r}_h = \left(1-\beta\right) {G M_\odot}/(r_h(t))^2$. Hence dust and nucleus move on Keplerian trajectories, modulated by $\beta$.
    \item[D3.]  We assume that the initial velocity of the nucleus, $\Vec{v_n(t_0)}$, is parallel to the $\eta$-axis. This implies that our model nucleus is, for the values used, at perihelion at the time $t_0$.
    \item[D4.] As the approximation for the relevant forces described in Section~\ref{sec:forces} is only applicable at a sufficient distance from the comet, the dust trajectories cannot be studied near the comet's surface. Instead, in our model the dust particles originate on a spherical surface with $r=\SI{1e4}{\meter}$ around the comet. The starting points are distributed randomly across this surface.
    \item[D5.] The initial velocity of a dust particle $i$ emitted at time $t_{em}$, $\Vec{v^i_{dust}}(t_{em})$, is described as the sum of the nucleus velocity at time $t_{em}$ and an assigned component $\Delta \Vec{v^i_{dust}}$: $\Vec{v^i_{dust}(t_{em})} = \Vec{v_n(t_{em})} + \Delta \Vec{v^i_{dust}}$. The starting velocity relative to the comet $\Delta \Vec{v^i_{dust}}$ depends on the radius of the dust particles and points radially outward from the comet. For the speed we use values taken from \citet{Marschall2020}, which predicts a dependence of $\Delta v^i_{dust} \propto \frac{1}{\sqrt{a^i}}$.
\end{enumerate}

For the comet's initial position and velocity we use $r_h(t_0) = \xi_0 = \SI{2.19} \AU$ and $v_\eta(t_0) = \SI{2.34e4}{\meter\per\second}$. These roughly correspond to the heliocentric distance and speed of comet 67P on 1st March 2015 as obtained from the JPL Horizons system\footnote{https://ssd.jpl.nasa.gov/horizons/app.html\#/}. 
At this point in time, the true angle between the heliocentric position and velocity vectors is 124$^\circ$, while we assume them to be perpendicular (D3) to allow for an easier implementation. Since we do not intend to simulate the exact observational circumstances of every data set investigated by \cite{Bertini2017}, but rather concentrate on one representative case, we consider our simplification (D3) justifiable. In the appendix, we show that the exact choice of the initial position and velocity of the comet have no impact on the observed results. Our simulations typically cover $\approx$46 days (see below).

\begin{figure}
    \centering
    \includegraphics[width=\linewidth]{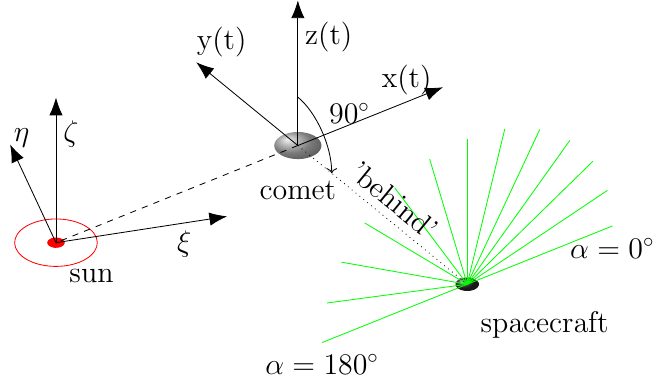}
    \caption{Observation geometry for the constellation 'behind'. The green lines correspond to the different cones of vision. }
    \label{fig:obsgeom}
\end{figure}

After calculating all trajectories in the heliocentric, non-rotating, inertial coordinate system, these solutions are transformed into a comet-centered, Cartesian frame of reference (see \autoref{fig:obsgeom}). This coordinate system moves with the position of the comet and is rotating in such a way, that the x-axis always points away from the sun and the y-axis is perpendicular to the x-axis and lies in the plane of movement of the comet. The z-axis completes the Cartesian system.

\subsection{Numerical implementation}

The relative position of a dust particle and the nucleus mainly depends on the time since emission, $\Delta t$, the assigned ejection velocity $\Delta \Vec{v^i_{dust}}$, and $\beta$, but only weakly on the absolute time of emission $t_{em}$. This is immediately obvious if we consider the comet to be on a circular orbit, with particles being emitted in regular time steps. Due to symmetry reasons, the trajectories of these particles have to be identical in the comet's frame of reference.

For our model, we can write in the rotating frame of reference:

\begin{align}
    \Delta \Vec{x}(\Delta t) &= \Vec{x_p}(t_{em}+\Delta t) - \Vec{x_n}(t_{em}+\Delta t) \\ \nonumber
    &= \Vec{x_p}(t_{em}) + \Vec{v_p}(t_{em})\Delta t + \int_{t_{em}}^{t_{em}+\Delta t}  \int_{t_{em}}^{t'} \Vec{a_p}(\tau) d\tau dt' \\ \nonumber
    &\quad - \left(\Vec{x_n}(t_{em}) + \Vec{v_n}(t_{em})\Delta t + \int_{t_{em}}^{t_{em}+\Delta t}  \int_{t_{em}}^{t'} \Vec{a_n}(\tau) d\tau dt'\right) \\ \nonumber
    &= \Delta \Vec{x}(t_{em}) + \Delta \Vec{v}(t_{em}) \Delta t + \int_{t_{em}}^{t_{em}+\Delta t} \int_{t_{em}}^{t'} (\Vec{a_p}(\tau) - \Vec{a_n} (\tau)) d\tau dt' \\ \nonumber
\end{align}
with the position vectors $\Vec{x}$, the velocities $\Vec{v}$ and accelerations $\Vec{a}$, where $p$ and $n$ denote the particle and nucleus respectively. In this notation, the accelerations $\Vec{a}$ include Coriolis forces and are thus not only dependent on the $\beta$-values and heliocentric distances, but also the velocities of the corresponding particles.

In the co-rotating reference frame, the assumptions leading to the choice of $\Delta\Vec{x}(t_{em})$ and $\Delta\Vec{v}(t_{em})$ for a given particle are independent of $t_{em}$ (D4,D5). The only dependence on $t_{em}$ is contained in the accelerations, because they depend on the comet's heliocentric distance and velocity relative to the sun which change over time for a non-circular orbit. During the typically covered integration time, the comet's heliocentric distance in our simulation increases by $\approx1.1\%$ and its velocity decreases by $\approx0.8\%$. At maximum, the velocity of the comet deviates from the y-direction by $\SI{4.2}{\degree}$. We assume these changes to be small enough to consider the orbit as circular for the purpose of treating the relative position of a dust particle and the nucleus as independent of $t_{em}$.

Thus, instead of simulating the trajectories of particles emitted at various emission times, we start the emission of all dust particles at time $t_0$ and interpret each calculated trajectory as a series of dust particles released at different times. This allows us to simulate the entire coma with only a fraction of the computational effort. We have verified numerically that our results do not deviate significantly from the results for a circular orbit and are in broad agreement with results from a simulation code that does not make this simplification (see \autoref{fig:circ_orb}).

For the calculation of the trajectories of dust particles and comet we used the solve\_ivp routine of the scipy.integrate python package. This returns not only the final positions of each object, but also its position in predefined time steps.


In order to achieve a sufficient resolution, most simulations use $N = \SI{3e5}{}$ different starting points and $n = \SI{1e5}{}$ particles along each trajectory, i.e. time steps. The trajectories are calculated up to a maximum time $t_{max}$, for most simulations this value is set to $t_{max} = \SI{4e6}{\second} \approx \SI{46}{\day}$. Unless mentioned otherwise, simulations were calculated using these parameter values.

\begin{figure}
    \centering
    \begin{subfigure}[b]{\linewidth}
        \includegraphics[width=\linewidth]{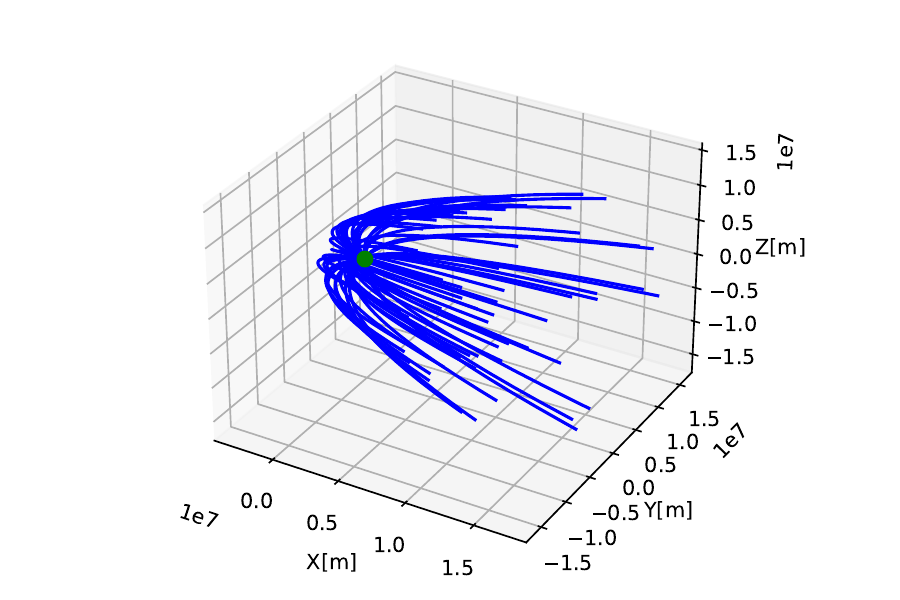}
        \caption{$t_{max} = \SI{5e5}{\second}$}
        \label{fig:trajvis_sf1}
    \end{subfigure}
    \hfill
    \begin{subfigure}[b]{\linewidth}
        \includegraphics[width=\linewidth]{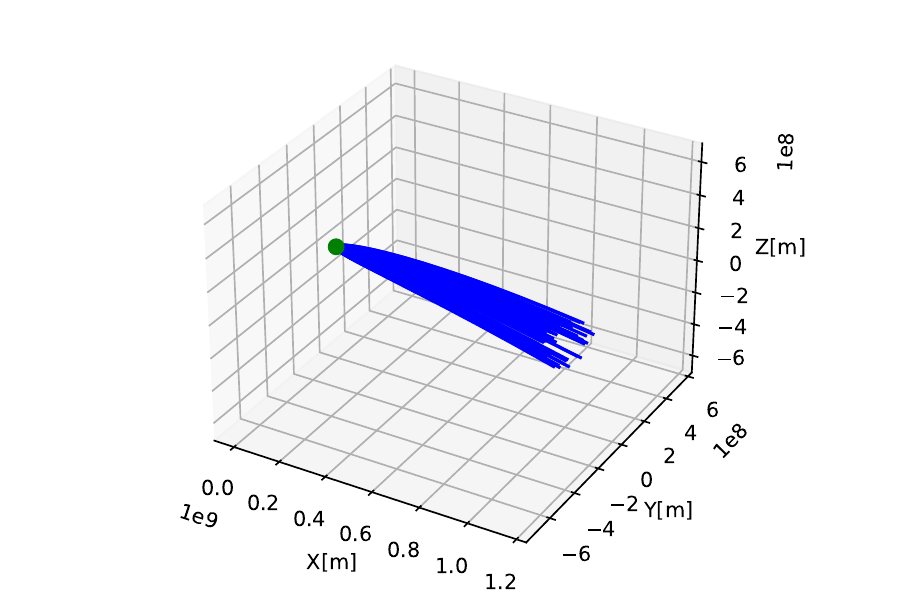}
        \caption{$t_{max} = \SI{4e6}{\second}$}
        \label{fig:trajvis_sf2}
    \end{subfigure}
    \caption{Calculated trajectories displayed in the rotating frame of reference of the comet. For visualization purposes we only display $N = 64$ starting points and restrict the starting points to the day side of the comet. The green dot represents the comet and is not to scale.}
    \label{fig:trajvis}
\end{figure}

The resulting trajectories in the comet-centered frame of reference from two such simulations are depicted in \autoref{fig:trajvis}, the difference between the simulations being the maximum integration time. This shows, that the model produces density structures which are at least qualitatively in alignment with our expectations. Dust particles emitted directly at the subsolar point move around $\SI{3.2e6}{\meter}$ towards the sun, before they turn around and move into the tail. This turnaround takes place roughly $\SI{2.1e5}{\second}$ after ejection.

\subsection{Activity distribution}

In order to simulate a non-uniform distribution of dust emission across the surface of the comet, each particle, $i$, is assigned a weight value $w_i$ according to its starting position. This weight is equivalent to the relative activity at the starting position, meaning particles in high-activity regions have a high weight value. For most activity distributions we use, the activity on the night side is zero. The weight distributions are normalized to represent the same mass flux across the whole sphere, meaning that the surface integral across the sphere from which the dust particles originate
\begin{equation}
    \int_0^{2\pi} \int_0^ \pi \sin{\theta} \cdot \omega\left(\theta,\phi\right) \: d\theta d\phi = \text{const.}
\end{equation}
has the same value for all weight distributions. Due to the same starting positions, dust particles along a given trajectory all have the same assigned weight $w_i$.

For most simulations, we use a cosine activity distribution across the dayside:
\begin{equation}
    w_{cos} = \begin{cases}
     \cos{\gamma} & \text{on the day side}\\
     0 & \text{on the night side},
   \end{cases}
\end{equation}
where $\gamma$ is the 3D-angle between the starting position of the particle, the comet center and the subsolar point ($\SI{0}{\degree}$ at the subsolar point, $\SI{90}{\degree}$ at the day-night-border). We refer to this distribution as the cosine distribution.

To a lesser extent, we also use a constant activity across only the day side
\begin{equation}
    w_{eq} = \begin{cases}
     \frac{1}{2} & \text{on the day side}\\
     0 & \text{on the night side},
   \end{cases}
\end{equation}
a uniform distribution across the whole sphere
\begin{equation}
    w_{un} = \frac{1}{4},
\end{equation}
a combined distribution with a constant (low) activity on the night side and a modified cosine distribution on the day side
\begin{equation}
    w_{comb} = E \cdot \begin{cases}
     max(\cos(\gamma), e) & \text{on the day side}\\
     e & \text{on the night side},
   \end{cases}
\end{equation}
with the normalization constant
\begin{equation}
    E = \frac{1}{1+2e+e^2}
\end{equation}
and a typical value of $e = 0.1$. For $e = 0$ this is identical to the cosine distribution. Finally, we study a squared cosine distribution across the whole sphere:
\begin{equation}
    w_{\cos^2} = \frac{1}{2} \cos^2\frac{\gamma}{2}.
\end{equation}

We note that each starting point is fixed with regard to its angular position to the sun and is not  affected by the rotation of the comet.

\subsection{Simulation of Rosetta/OSIRIS measurements}

To match our density data to the observed data by \citet{Bertini2017}, we need to evaluate the column area density in a given cone of vision from Rosetta, analog to the measurements evaluated by \citet{Bertini2017}. By assuming that the coma is optically thin and the particles are in each other's far zone multiple scattering can be neglected. In such a case, the flux received by the detector can be written as 
\begin{equation}
F = F^{\rm inc} \sum_{i} w_i A_{i} Q_{\rm sca, i} \frac{P_{i}}{4\pi}\frac{1}{d_i^2},
\end{equation}
where $F^{\rm inc}$ is the incident flux, $A_i$ is the cross section, $Q_{\rm sca, i}$ is the scattering efficiency, $P_{i}$ is the phase function, $w_i$ is the assigned weight, and $d_i$ is the distance from the detector of the $i$:th particle. The summation is over all the particles inside a given cone of vision.

As the scattering properties, i.e., the single scattering phase functions $P_i$ and the scattering efficiencies $Q_{\rm sca}$ of dust particles are unknown,  we cannot match the measured radiance profiles directly to the model. We can, however, define the column area density which neglects all the scattering properties of the particles. This quantity describes the effects of the phase angle dependent column area density on the measured radiance profile, and is given by
\begin{equation}
    \rho_{opt} = \sum_i w_i \frac{A_i}{d_i^2}.
    \label{eq:colareadens}
\end{equation}
Consequently, the measured radiance profiles by \citet{Bertini2017} are combinations of the dust phase function and the column area density.

With the cross-section of a particle $A \propto a^2$, the column area density (disregarding all prefixes and for a single particle size) is
\begin{equation}
    \rho_{opt} = a^2 \sum_i \frac{w_i}{d_i^2}.
\end{equation}
We assume that the column area density is sufficiently small, so that no dust particles are obscured by others. Also, since we are only interested in the general shape of the function, the used units for the column area density are arbitrary. This dimensionless quantity can be understood as the filling factor of the dust particles in a given field of view, describing the solid angle fraction of a pixel covered by dust.

For our simulated observations it is assumed that Rosetta is not moving in the rotating, comet-centered frame of reference. For most simulations Rosetta trails the comet along its trajectory at a distance of $\SI{2e5}{\meter}$, i.e. lies in negative y-direction, see \autoref{fig:obsgeom}.

The cones of vision are in a plane perpendicular to the plane defined by Sun, Comet and Rosetta. The observational geometry for a typical constellation is depicted in \autoref{fig:obsgeom}.

Each cone has an opening angle of $\SI{2.2}{\degree}$, which corresponds to the field of view of the camera used by \citet{Bertini2017}. We evaluate cones at 45 evenly spaced phase angles between $\SI{0}{\degree}$ (looking away from the sun) and $\SI{180}{\degree}$ (looking towards the sun).

\section{Results}

\subsection{Relevance of the spacecraft position}

\begin{figure}
    \centering
    \includegraphics[width=\linewidth]{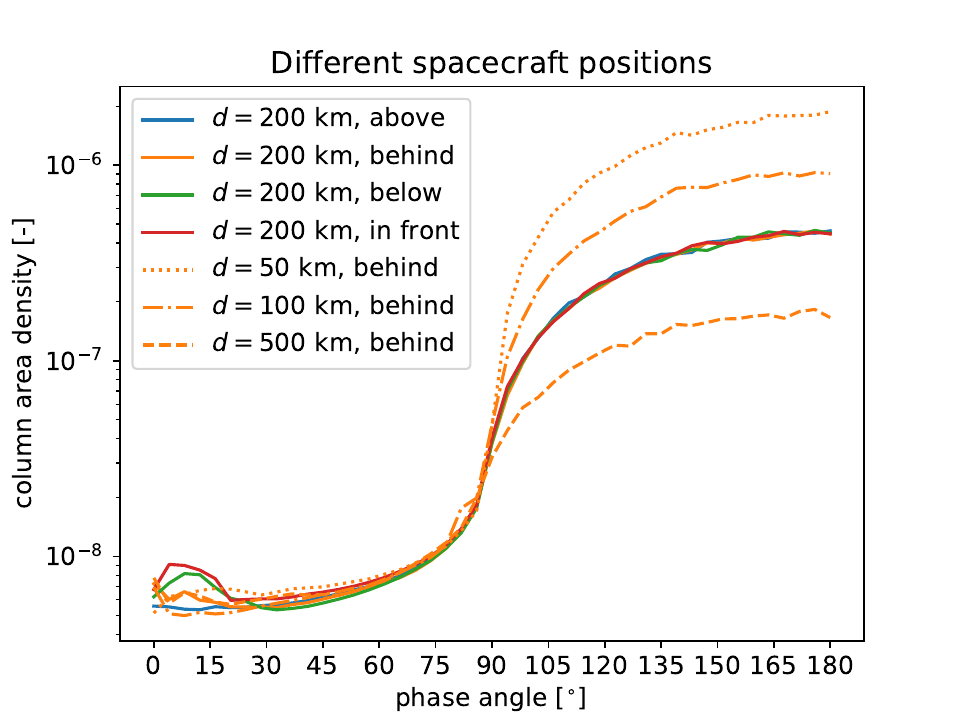}
    \caption{Influence of different spacecraft positions on the measured column area density. The default "behind" position is shown in \autoref{fig:obsgeom}. The used dust parameters are $a = \SI{1e-5}{\meter}, v_{dust} = \SI{3e1}{\meter\per\second}$, the activity is $w = \cos{\gamma}$.}
    \label{fig:position_var}
\end{figure}

The measurements used by \citet{Bertini2017} were taken at very different distances between comet and spacecraft, ranging between $\SI{3.5e4}{\meter}$ and $\SI{1.2e6}{\meter}$. In spite of these varying conditions, the measured phase functions show no systematic difference.

In a first step we want to analyze the impact of the position of the spacecraft on the measurement. \autoref{fig:position_var} shows the column area density as observed from different spacecraft positions as a function of the phase angle. We vary the distance between spacecraft and comet, as well as the angular position of the spacecraft ("in front", "behind": positive, negative y-direction, "above", "below": positive, negative z-direction, see \autoref{fig:obsgeom}). For the activity distribution we used the cosine distribution.

\autoref{fig:position_var} shows that the overall dependence of the column area density on the phase angle does not change with the spacecraft position. In all cases, the column area density reaches plateaus at low and high phase angles, with the column area density at high phase angles being about two orders of magnitude larger. The increase of the column area density takes place between $\SIrange{45}{135}{\degree}$ and appears to be symmetric around $\SI{90}{\degree}$.

We observe an increase of the column area density at high phase angles with decreasing distance to the comet. This is plausible, as the high phase angle column area density is dominated by young particles close to the spacecraft, which will be closer and thus have a larger impact on the column area density with decreasing distance between spacecraft and comet. As the spacecraft position does not affect the column area density at low phase angles, we can assume that this regime is dominated by particles far away in the tail.

At very low phase angles ($<\SI{30}{\degree}$) we observe a slight increase for some spacecraft positions. We suspect that this is a numerical artifact, which appears when a trajectory passes very close to the spacecraft through a cone of vision. As the impact on the column area density of a particle scales with $\frac{1}{d^2}$, these close particles dominate. This effect is most prevalent at low phase angles, where the column area density is low and mostly stems from particles in the tail.

For the following simulations the spacecraft is positioned at a distance of $d = \SI{2e5}{\meter}$ behind the comet.

\subsection{Influence of dust size and velocity}

\begin{figure}
    \centering
    \includegraphics[width=\linewidth]{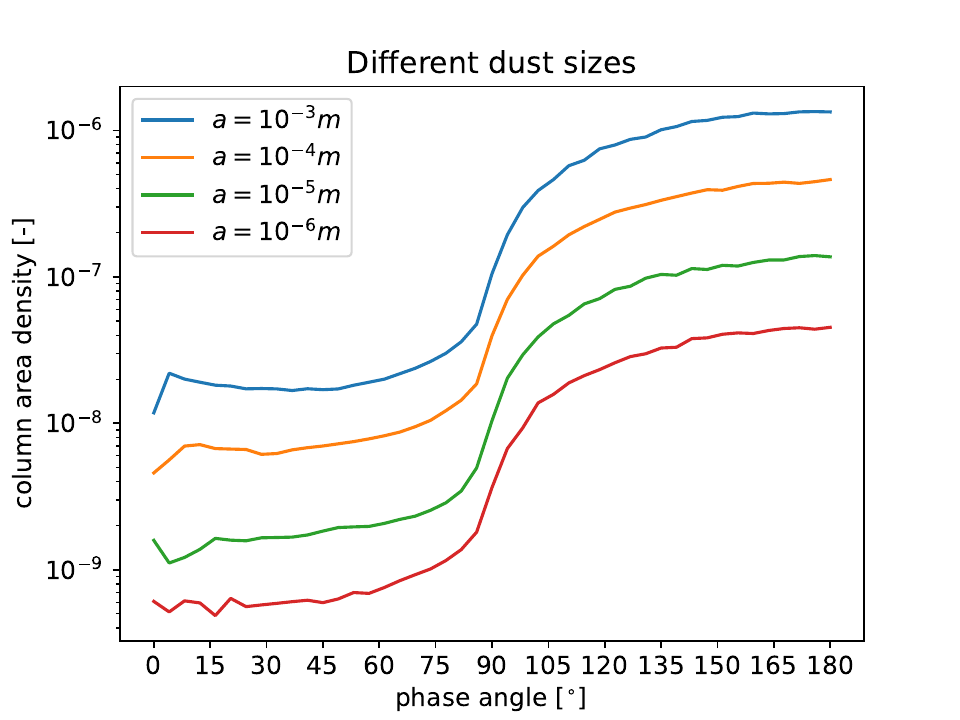}
    \caption{Influence of different particle sizes on the measured column area density. The activity distribution is $w = \cos{\gamma}$. The simulation with $a = \SI{1e-3}{\meter}$ uses a higher maximum integration time of $t_{max} = \SI{1.6e7}{\second}$. The simulations are normalized to represent the same mass flux.}
    \label{fig:dust_var}
\end{figure}

As the size of the dust particles leaving the comet is not fully known, we calculate the column area densities for different particle sizes: $a = \SI{1e-3}{\meter},\SI{1e-4}{\meter},\SI{1e-5}{\meter},\SI{1e-6}{\meter}$.

As mentioned in \autoref{sec:geom}, the starting velocity of dust particles depends on the particle size. In agreement with \citet{Marschall2020} we use (in the same order as above): $v_{dust} = 
\SI{3}{\meter\per\second}, \SI{10}{\meter\per\second}, \SI{30}{\meter\per\second}, \SI{100}{\meter\per\second}$.

\autoref{fig:dust_var} shows the column area density for the different particle radii and corresponding starting velocities. For the activity we once again use the cosine distribution. The results are normalized to represent the same mass flux, correcting for the mass increase with larger radius per particle.


As is immediately obvious, the column area density is higher for bigger particles. This is due to a smaller difference between the size of the comet and the dust and the resulting similarity in their $\beta$ values. As a result, the difference in acceleration between comet and dust is smaller and the dust remains longer in close proximity to the comet. Particles that are further away have a smaller apparent cross-section and thus a smaller impact on the column area density. This effect appears to be stronger than the decrease of surface area per volume and lower initial velocity with increasing particle radius.

A quick order of magnitude estimation confirms these results: The observed column area density should be proportional to the number density of emitted dust particles $N_\text{Dust}$, the cross-section of a single dust particle $A$ and the squared inverse of some mean distance $d$ of the dust particles to the observer:
\begin{equation}
    \rho_{opt} \propto N_\text{Dust} \cdot A \cdot \frac{1}{d^2}.
\end{equation}
Since we assume a constant mass flux, we know that $N_\text{Dust} \propto a^{-3}$. Trivially we know the cross-section of a spherical particle to be $A \propto a^2$. Assuming that the relative velocity between dust and comet is dominated by the difference in acceleration and not the initial velocity of the dust, we get
\begin{equation}
    d \propto v_\text{rel}
    \propto F_\text{rel}
    \propto \left|F_\text{Dust} - F_\text{67P}\right|
    \propto \left|\beta_\text{Dust} - \beta_\text{67P}\right|,
\end{equation}
where $v_\text{rel}, F_\text{rel}$ are the differences of the velocity, forces acting on the comet and the dust and $F_\text{Dust}, F_\text{67P}$ are the individual forces as defined by \autoref{eq:forces_0}.
Using that $\beta_\text{67P} << \beta_\text{Dust}$ and $\beta \propto a^{-1}$, we get
\begin{equation}
    d \propto \beta_\text{Dust} \propto a^{-1}.
\end{equation}
Putting all this together, we get
\begin{equation}
    \rho_{opt} \propto a,
\end{equation}
which, in qualitative agreement with the data, predicts an increase in column area density for larger particles.

The overall dependence of the column area density on the phase angle however does not change with the dust size. Similar to \autoref{fig:position_var} we observe small disturbances at low phase angles, otherwise there is no significant difference.

\subsection{Influence of the activity distribution}

In all simulations shown so far, we have used the cosine activity distribution. We now want to investigate how sensitive our model is to different activity distributions across the surface.

\begin{figure}
    \centering
    \includegraphics[width=\linewidth]{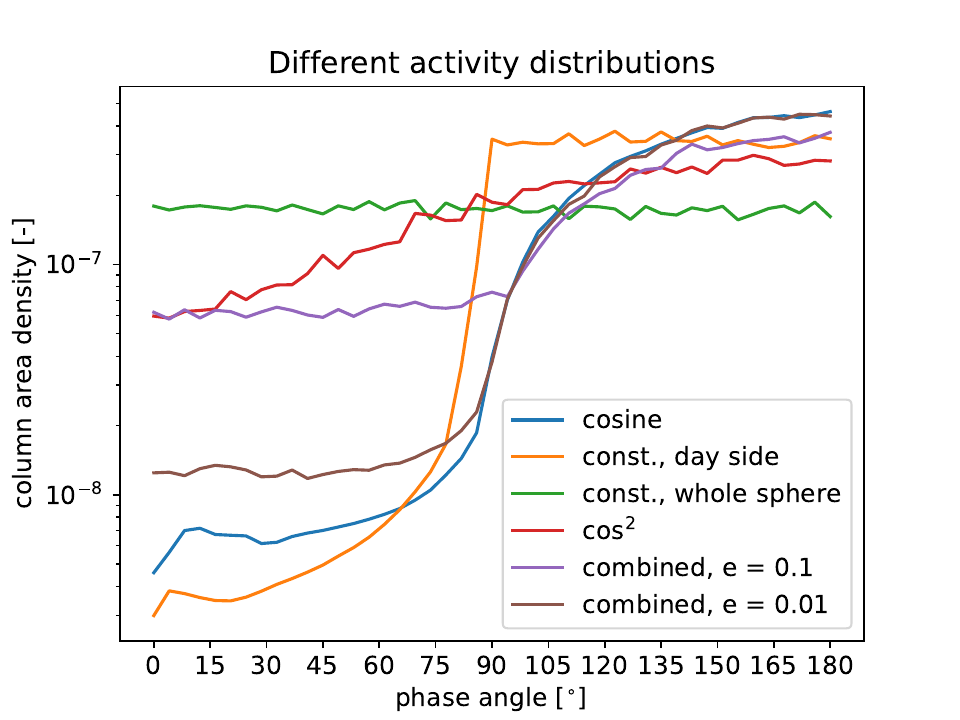}
    \caption{Influence of different surface activity distributions on the measured column area density. The dust radius is $a = \SI{1e-5}{\meter}$.}
    \label{fig:dist_var}
\end{figure}

\autoref{fig:dist_var} shows the column area density for the different activity distributions. Unlike the dust size and the position of the spacecraft, the surface distribution has a significant impact on the dependence of the column area density on the phase angle. This is most obvious for the constant activity across the whole comet, for which we observe a column area density that is mostly independent of the phase angle. Comparing this to the constant activity on the day side, we see a similar behavior at high phase angles (except for an offset by a factor $\approx 2$ due to the normalisation), with a steep decline to lower phase angles, similar to the cosine distribution. 

Interestingly, the constant distribution across the day side reaches lower column area densities at low phase angles compared to the cosine distribution. This suggests that particles starting near the terminator have a lower impact on the observed density in the tail compared to particles that start near the subsolar point. We find this to be plausible, because the starting velocity of particles near the terminator points perpendicular to the tail, meaning they will be in the outer edges of the tail.

The "combined" distributions show a flat column area density profile at low phase angles, similar to the constant activity distribution across the whole comet. At high phase angles they behave like the cosine distribution. The relative activity on the night side $e$ determines the density at low phase angles, with higher night side activities corresponding to higher densities.

For the squared cosine distribution we observe a weaker dependence on the phase angle. Also, we do not observe a drastic shift in behavior around $\SI{90}{\degree}$ as opposed to the other distributions. This suggests that such a shift is directly caused by the used activity distributions and not by the particle dynamics at the terminator.

The impact of the activity distribution on the measured column area density shows that particles close to the spacecraft dominate the column area density. The particles far away in the tail only gain relevance when there is no or next to no activity on the night side. Thus, the assumed surface activity distribution plays a major part in the expected column area density measurements.

\section{Conclusion}

We used a simple numerical model to calculate the dust density in the coma of 67P/Churyumov-Gerasimenko, in which the only acting forces are the solar gravity and radiation pressure. Using this density, we calculated the column area density as seen from Rosetta's position for a wide range of phase angles, similar to the measurements conducted by \citet{Bertini2017}. We investigated the relevance of the spacecraft position relative to the comet and conducted parameter studies over the dust size and the activity distribution across the surface of 67P/Churyumov-Gerasimenko.

We found the observed dependence of the column area density on the phase angle to be largely independent of the angular position of the spacecraft within the terminator plane. The distance between comet and spacecraft only affects the column area density at high phase angles, with higher distances corresponding to lower column area densities. The dust size only influences the column area densities by a constant factor, where larger particles stay longer in close proximity to the comet and the spacecraft and thus result in higher measurements for the column area density.

We determined the most important parameter to be the activity distribution across the surface of the comet. For models with no night side activity, we find the column area density at high phase angles roughly 2 orders of magnitude higher than at low phase angles. For distributions with included night side activity, this difference decreases depending on the activity level on the night side.

Our simulation results imply that the Rosetta/OSIRIS phase function \citep{Bertini2017} does not correspond to the single particle phase function for a realistic activity distribution. They are equivalent only for the isotropic activity distribution which is unlikely due to the day-night variation of the solar illumination \citep{Gerig2020}. Since the activity distribution is not well known, it is not trivial to constrain the dust properties from the Rosetta/OSIRIS phase function.

The model can be easily adjusted to the parameters of other comets and new insights into parameters like the dust size, bulk density and activity models. In the future, the here obtained results can be combined with measured intensity data by \citet{Bertini2017} to gain further insight on the scattering phase function of the dust particles.

\section*{Acknowledgements}
J.A.'s contribution was made in the framework of the European Research Council (ERC) Starting Grant No. 757390 (CAstRA). J.A. acknowledges funding by the Volkswagen Foundation. J.M. was supported by the German Research Foundation (DFG) grant no. 517146316.

\bibliographystyle{aa}
\bibliography{lit_astroprak.bib}

\begin{appendix} 

\section{Justification for the simplified Orbit model}

\renewcommand\thefigure{A.\arabic{figure}}
\begin{figure}[!htb]
    \centering
    \includegraphics[width=\linewidth]{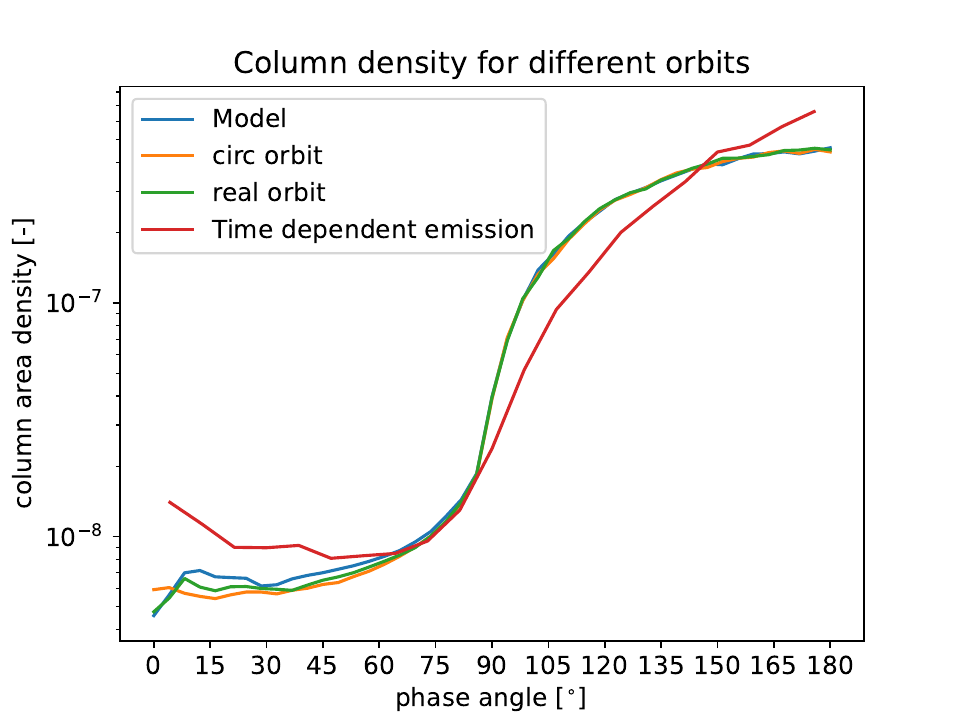}
    \caption{Comparison between a circular orbit, the real orbit, our selected model and a simulation calculated with an independent code that accounts for the time- and space-dependent dust emission. All four simulations use identical parameters except for the starting velocity of the comet, which was adjusted to produce the respective orbits. We use a cosine distribution with the dust parameters $a = \SI{1e-5}{\meter}$, $v = \SI{10}{\meter\per\second}$. The last simulation was generated using a Monte Carlo dust dynamics code that emits particles at different times along the comet's trajectory in which the dust production rate is proportional to the heliocentric distance as $r_h^{ -2}$.}
    \label{fig:circ_orb}
\end{figure}

\begin{figure}[!htb]
    \centering
    \includegraphics[width=\linewidth]{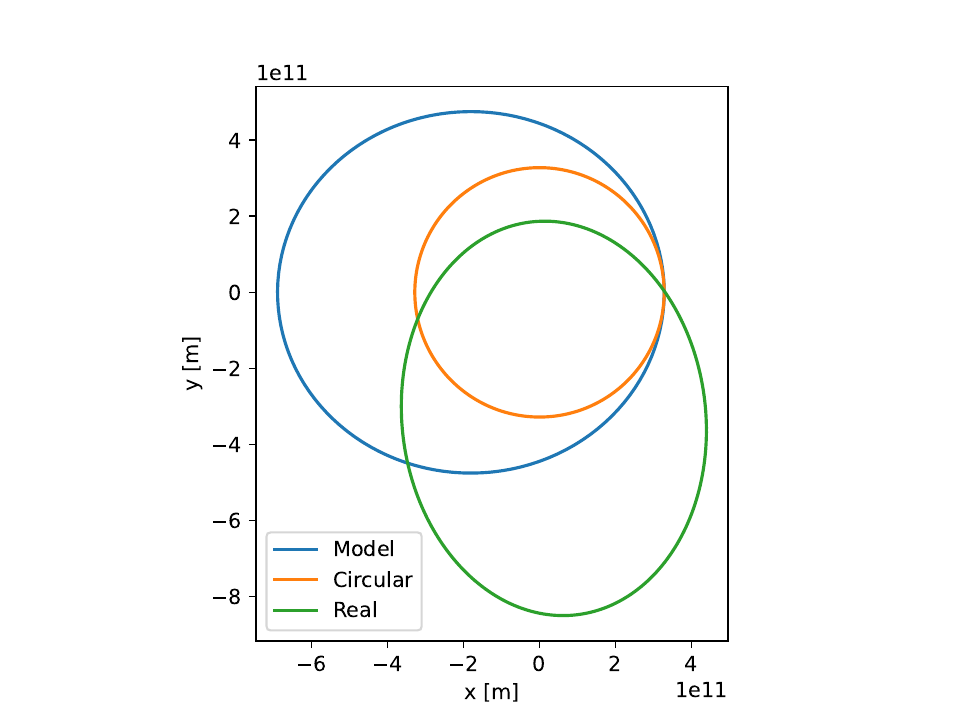}
    \caption{Comparison of the orbits for different starting velocities. The "real" orbit also has a marginal $z$-velocity, while the other two move exclusively in the x-y-plane.}
    \label{fig:orbs}
\end{figure}

\autoref{fig:circ_orb} shows the column area density for different orbits (shown in \autoref{fig:orbs}) as well as for a different model which does not neglect the dependence of the particle trajectories on the time of emission. We observe no significant difference between the three models using our simplification. The model using the real orbit of the comet 67P and time-dependent emission shows a minor difference due to the elliptical orbit. The general shape of the column area density, however, remains the same---two orders of magnitude difference between the forward and backward directions. The required computational time for the time-dependent emission model is significantly longer than for the simplified model.


\end{appendix}

\end{document}